# Controlling Atom-Surface Scattering with Laser Assisted Quantum Reflection

A. L. Harris[1]

[1] Physics Department, Illinois State University, Normal, IL, USA 61790

## Abstract

In low energy atom-surface scattering, it is possible for the atom to be reflected in a region of attractive potential with no classical turning point. This phenomenon has come to be known as quantum reflection and it can reduce the sticking probability of atoms to surfaces, as well be used for atom trapping. We simulate the quantum reflection process in a one-dimensional model with a slow-moving atom moving in a Morse potential in the presence of an applied laser field. We show that in the case of laser-assisted quantum reflection, the laser field imparts additional momentum and kinetic energy to the atom. This results in a decreased distance of closest approach between the atom and surface. Our results show that the distance of closest approach and can be controlled through the timing and intensity of the laser pulse, which may result in enhanced sticking probability and/or reduced quantum reflection probability.

## 1. Introduction

In atom-surface scattering, there is a well-known quantum phenomenon often referred to as quantum reflection. Near-threshold, low energy atoms approaching a surface can be reflected in a region where the interaction potential is attractive and there is no classical turning point. As the kinetic energy of the atom approaches zero, the reflection probability tends to one and the point of reflection occurs at a larger spatial distance from the surface. This phenomenon has been experimentally observed in several physical systems. Specifically, the sticking probability of atoms incident on liquid helium was shown to be reduced due to quantum reflection [1–6]. Quantum reflection has also been observed for neon atoms impinging on solids [7–9], as well as in cold atom collisions and the interaction of Bose-Einstein condensates with surfaces [10–12], even being proposed as a means for trapping cold atoms [13,14].

From a theoretical standpoint, quantum reflection is nothing new. One textbook example is that of a particle colliding with a step potential. When the energy of the particle is above that of the step, both reflection and transmission occur. This is a simple case of quantum reflection, where the abrupt change in potential causes a rapid change in wave number resulting in reflection that can only be explained through the wave nature of matter. What makes atom-surface quantum reflection different is that it occurs in a region where the potential is purely attractive, and one would not expect to see reflection. In fact, the reflection point for low energy atoms incident on surfaces can occur at tens or hundreds of nanometers above the surface, well before the atom can adhere to the surface [15].

The possible applications of quantum reflection are many and range from atomic mirrors [7,12,16] to matter wave diffraction gratings [17] to methods for trapping and controlling cold atoms [13,18–21]. In order to realize some of these applications, efforts are underway to design and engineer materials and devices that use quantum reflection for the control and manipulation of matter waves. For example, in [13,22], a method was presented for suppressing quantum reflection through electrical control of an applied voltage to a grating. This process altered the interaction potential between the atom and the surface such that quantum reflection was limited. In [23], quantum effects were enhanced by altering the interaction potential to include sharp features, and in [24] a magnetic field was used to control quantum reflection in graphene.

We propose an alternate mechanism by which quantum reflection can be controlled, namely laser-assisted quantum reflection. We numerically solve the one-dimensional Schrödinger equation for a slow atom approaching a reflective surface. At various times prior to the atom's reflection a limited duration external electric field is applied causing the atom's trajectory, momentum, and energy to be altered. Following the application of the laser field, the atom is allowed to propagate freely in the presence of the surface potential. Our results demonstrate that the introduction of the laser pulse at certain times during the quantum reflection process can change the location at which quantum reflection occurs. We also demonstrate that control of the distance of closest approach of the atom to the surface can be controlled through the intensity of the applied laser field. The combination of controlling both the timing and intensity of the laser pulse offer a means to control the quantum reflection process, thereby possibly suppressing quantum reflection and/or enhancing adsorption of the atom to the surface. Atomic units are used throughout unless otherwise noted.

## 2. Theory

To simulate quantum reflection for a slow atom impinging on a surface, we solve the one-dimensional Schrödinger equation [25]

$$i\frac{\partial}{\partial t}\psi(x,t) = \left[-\frac{1}{2}\frac{d^2}{dx^2} + V(x) + xE(t)\right]\psi(x,t), \qquad (1)$$

where $V(x)$ is the Morse potential

$$V(x) = V_0\left[\exp\left(-\frac{2(x-\alpha)}{d}\right) - 2\exp(-(x-\alpha)/d)\right]. \qquad (2)$$

The potential has a depth of $V_0$ and a minimum located at $x = \alpha$. The stiffness parameter is $1/d$. We use the parameters $V_0 = 1$ a.u., $d = 1$ a.u., and $\alpha = 0$. A plot of this Morse potential is shown in Fig. 1.

In the case of laser assisted quantum reflection, $E(t)$ is non-zero. For simulations without the laser field, $E(t) = 0$.

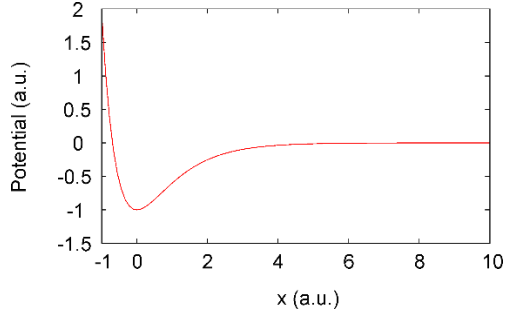

Figure 1 Morse potential with $V_0 = 1$ a.u., $d = 1$ a.u., $x_0 = 0$.

The incident atom is assumed to be a coherent Gaussian wave packet with average initial momentum $p_0$, average initial position $x_0$, and the standard deviation is $1/\sqrt{\Gamma}$

$$\psi(x, t = 0) = \left(\frac{\Gamma}{\pi}\right)^{1/4} \exp[-\Gamma(x - x_0)^2/2] \exp[-ip_0(x - x_0)]. \tag{3}$$

Note that $p_0 > 0$ and the initial wave packet is moving to the left.

The laser field is also assumed to have a Gaussian envelope given by

$$E(t) = E_0 e^{-2\ln 2 \left(\frac{t-t_L}{\Delta t}\right)^2} \sin(\omega_0(t - t_L)) \tag{4}$$

where $\omega_0$ is the laser frequency, $t_L$ is the temporal center of the laser pulse, and $\Delta t$ is the full width half maximum (FWHM) of the laser pulse. Because the oscillations of the laser field are much faster than the time scale of the quantum reflection process, we assume that the atom "feels" only the average of the field. Thus, it is the shape of the envelope function that is important in the laser assisted quantum reflection process.

## 3. Results

We propagate an initial Gaussian wave packet with $\Gamma = 10^{-10}$ a.u., $x_0 = 10^6$ a.u., and $p_0 = 10^{-4}$ a.u. Simulations were performed with and without the laser field. For simulations with the laser field, the field envelope had a full width half maximum of 5 x $10^8$ a.u. Three different laser field intensities of $I = 0.1, 10,$ and $1000 \, W/cm^2$ and three temporal centers of $t_L = 2, 4,$ and $6 \, \times 10^9$ a.u. were used.

Figures 2a and b show the temporal and spatial propagation of the wave packet without the laser field. In Fig. 2a, the complete time course of the quantum reflection process is shown from $t = 0$ to $t = 2 \times 10^{10}$ a.u. The reflection occurs at $t = 1 \times 10^{10}$ a.u. and interference structures are clearly visible in the wave function density. These structures are due to interference between the incoming and outgoing wave functions [26]. A close-up plot of the wave function density is shown in Fig. 2b. Here, the interference structures are more pronounced, and it is clear that the wave packet reflects at a distance well away from the minimum of the potential. The closest approach of the atom (as defined by the spatial distance of the maximum in the wave function density from zero) for no laser field is 15,750 a.u. (~830 nm).

Figures 2c-h show the temporal and spatial propagation of the wave packet in the presence of a $I = 10^3 \, W/cm^2$ laser field with a FWHM = $5 \times 10^8$ a.u. The center of the laser pulse was varied from $2 \times 10^9$ a.u. to $6 \times 10^9$ a.u. In all cases, the effect of the laser field is to hold the atom at a nearly constant distance from the surface while the field is present. The laser field imparts additional energy to the incident wave packet, and when the laser field is turned off, the atom resumes its motion toward the surface with increased velocity. This is evident from the decreased slope of the trajectory after the laser pulse compared to before the laser pulse, as shown Figs. 2b, d, and f.

The close-up plots in Figs. 2c, e, and g show that when a laser field is applied, the distance of closest approach is reduced, regardless of the time of pulse application. This is a direct consequence of the increased kinetic energy of the atom. The later the laser pulse is applied, the closer the atom approaches the surface. In the case of the latest applied pulse ($t_L =$ 6 x $10^9$ a.u.), the distance of closest approach is approximately 1800 a.u. This is a factor of nearly ten times closer than when no laser field is present (15,750 a.u.). In all cases, when the laser field turns off, the atom is still well beyond the region where the potential well is appreciably non-zero. The increased kinetic energy of the atom following the application of the laser field reduces its wavelength, which also contributes to its closer approach. This change in wavelength is observed in the interference pattern of the incoming and outgoing waves, as seen in Figs. 2b, d, f, and h. The increased kinetic energy of the incident atom following application of the laser field also reduces the time of interaction between the atom and the surface, as observed in the decreased temporal width of the wave function density at the atom's closest approach (Figs. 2b, d, f, and h).

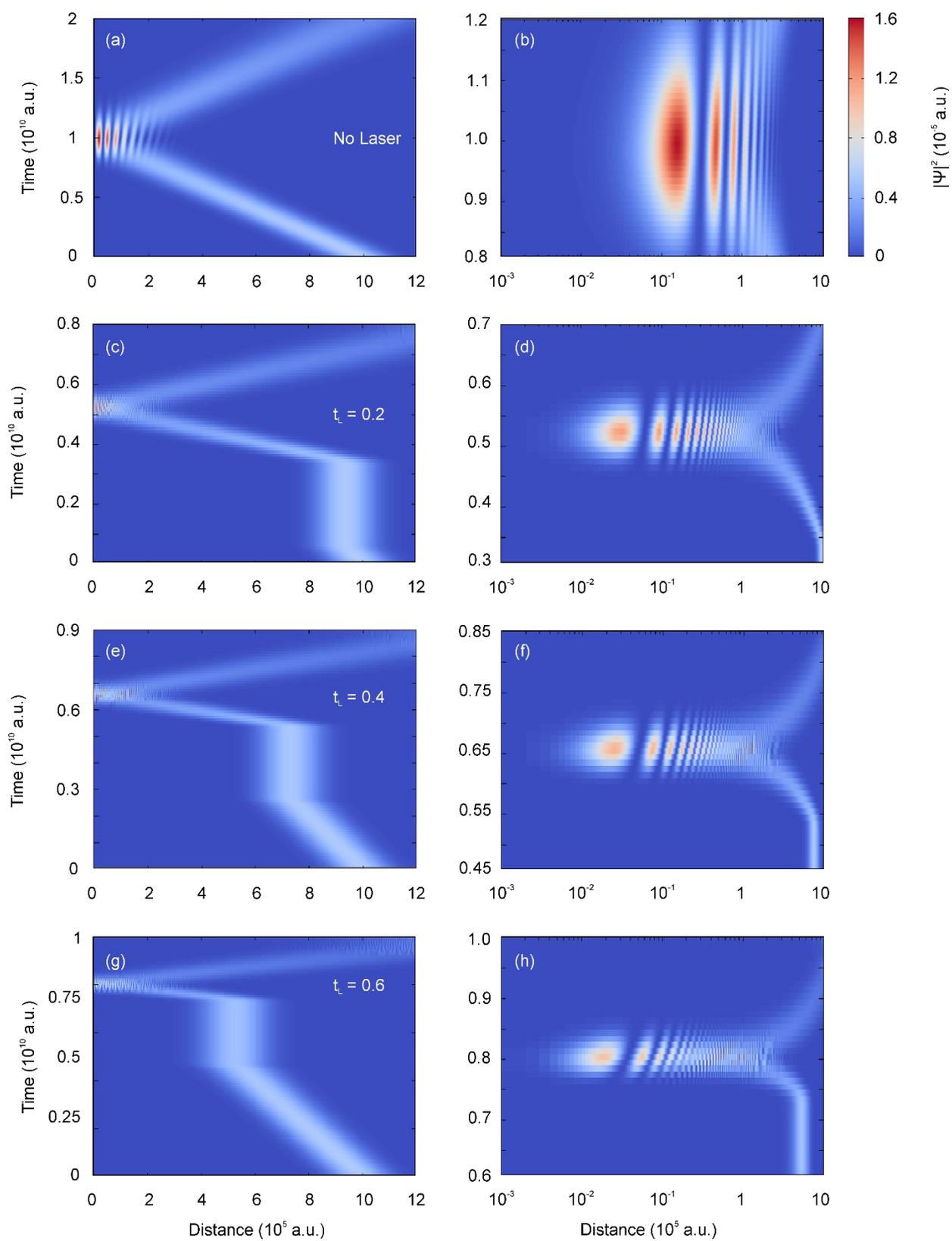

Figure 2 Spatial and temporal evolution of a Gaussian wave packet reflecting from a Morse potential. The incident momentum was $10^{-4}$ a.u. The left column (a, c, e, g) shows the full reflection process from the beginning of the simulation to after reflection is complete. The right column (b, d, f, h) shows a close-up view of the reflection as the wave packet nears the surface. The top row (a, b) shows model data without an external electric field. Panels (c-h) show model data for laser-assisted quantum reflection for a $10^3 \ W/cm^2$ pulse with a Gaussian envelope with FWHM = 5 x $10^8$ a.u. The temporal center of the pulse $t_L$ is listed in each panel.

Because the presence of the laser field alters the incident atom's kinetic energy, it should be possible to control the atom's distance of closest approach through the laser pulse's intensity. Figure 3 shows the temporal and spatial propagation of the wave packet with varying intensities of the laser field for a pulse centered at $t_L = 2$ x $10^9$ a.u. Despite varying the intensity over 4 orders of magnitude, only minor changes are observed in the spatial and temporal evolution of the wave packet.

As the intensity increases, the laser pulse is in effect for a long duration and the atom's trajectory is altered at an earlier time. This is observable in Figs. 3a, c, and e. The length of the vertical stripe in the density plot when the laser field is present is increased for larger intensities, indicating a longer effective pulse application time. This vertical stripe also begins at an earlier time for larger laser intensities, which alters the atom's trajectory at earlier times. The close-up plots of the reflection in Figs. 3b, d, and f show that the distance of closest approach decreases as laser intensity decreases. This effect is counterintuitive because a higher intensity laser imparts greater energy to the incident atom, which could result in a closer approach. Additionally, an atom with greater kinetic energy will have a smaller wavelength and the interference between the incoming and outgoing wave functions should occur at distances closer to the surface with a smaller separation between the interference maxima. However, the results of the simulation in Fig. 3 show the opposite. The lowest intensity laser pulse results in the closest approach of the atom and the smallest separation between interference maxima. While the Morse potential is attractive at large distances, the laser pulse adds a repulsive potential. This is what causes the atom to maintain a nearly constant position while the laser pulse is present. The added repulsive potential of the laser pulse decreases with decreasing intensity, thereby allowing atoms in a weaker laser field to more closely approach the surface.

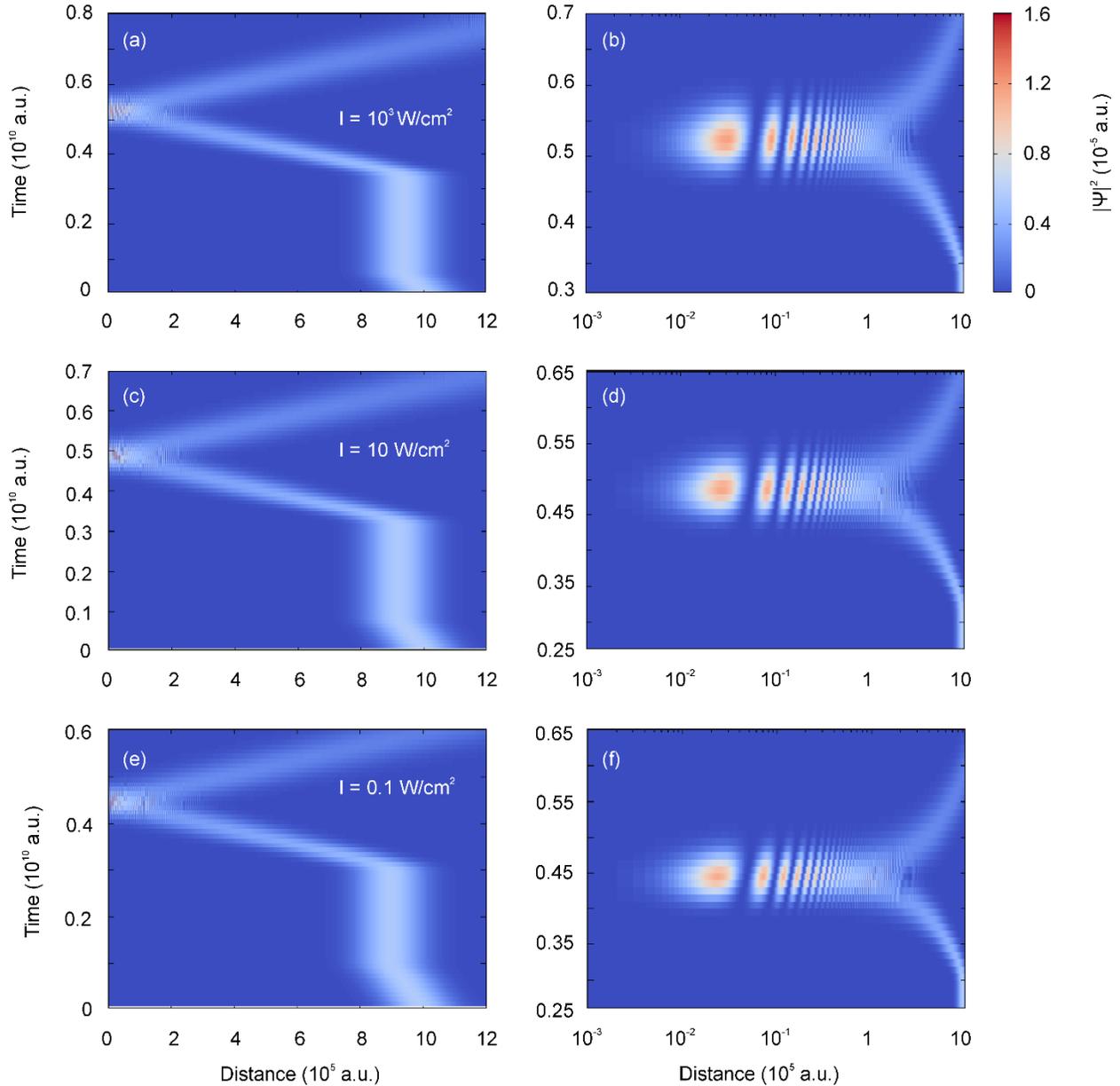

Figure 3  Spatial and temporal evolution of a Gaussian wave packet reflecting from a Morse potential in the presence of a laser field. The incident momentum was $10^{-4}$ a.u. The left column (a, c, e) shows the full reflection process from the beginning of the simulation to after reflection is complete. The right column (b, d, f) shows a close-up view of the reflection as the wave packet is near the surface. The laser intensities are labeled in the figure and a Gaussian envelope with FWHM = 5 x $10^8$ a.u. was used. The temporal center of the pulse was 2 x $10^9$ a.u.

Figure 4a shows the distance of closest approach of the atom for the different laser pulse application times and intensities.  Regardless of the time of application of the laser pulse, the distance of closest approach decreases with decreased intensity, indicating that the lowest intensity laser allows the atom to more closely approach the surface.  The distance of closest approach is smaller for later applied laser pulses, indicating that the laser pulse increases the kinetic energy of the atom most when applied to an atom is closer to the surface.

Figure 4b shows the average momentum of the atom after reflection for different pulse application times and intensities.  Because there is no absorption with the Morse potential, the final momentum is also the incident momentum prior to reflection.  Consistent with the data for closest approach, a lower pulse intensity results in a greater reflected momentum, which indicates a larger increase in kinetic energy of the atom due to the laser field.  Additionally, pulses applied when the atom is closest to the surface (later in time) result in a larger reflected atom momentum.  As discussed above, this increased kinetic energy results in smaller distances of closest approach.

The presence of a laser field during the quantum reflection process alters the incident atom's momentum and energy, resulting in the atom more closely approaching the surface.  This reduced distance of closest approach could enhance the atom's likelihood of adsorbing to the surface and reduce the quantum reflection probability.  The time and intensity of the applied laser field altered the distance of closest approach, which provides a mechanism for controlling the quantum reflection process.

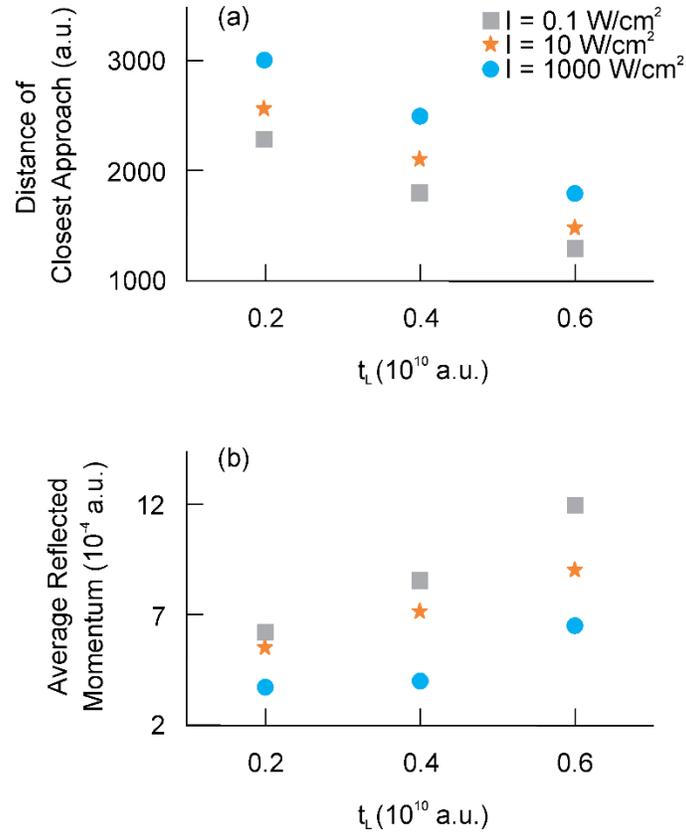

Figure 4 (a) Distance of closest approach of the wave packet to the surface for different temporal centers $t_L$ and intensities of the laser pulse. (b) Average final reflected momentum of the wave packet after interaction with the surface for different temporal centers $t_L$ of the laser pulse. The laser intensities are labeled in (a).

## 4. Conclusions

The quantum reflection process is a well-known quantum phenomenon in atom-surface scattering. Interference between the incident and reflected waves cause the atom to reflect in a region of a purely attractive potential. We performed simulations by solving the time-dependent Schrödinger equation and examined the effect of an applied laser field on the atom's distance of closest approach to the surface. We showed that the atom's distance of closest approach can be controlled through laser-assisted quantum reflection and that the presence of the laser field reduces the distance of closest approach. The timing and intensity of the laser pulse can be used to control the distance of closest approach. Later and less intense applied pulses result in smaller distances of closest approach, providing a potential mechanism to control adsorption and quantum reflection probabilities.

**Acknowledgements**
We gratefully acknowledge the support of the NSF under Grant No. PHY-1912093 and Illinois State University High Performance Computing resources.

References

[1] I. A. Yu, J. M. Doyle, J. C. Sandberg, C. L. Cesar, D. Kleppner, and T. J. Greytak, *Evidence for Universal Quantum Reflection of Hydrogen from Liquid $^{4}\mathrm{He}$*, Phys. Rev. Lett. **71**, 1589 (1993).
[2] V. U. Nayak, D. O. Edwards, and N. Masuhara, *Scattering of $^{4}\mathrm{He}$ Atoms Grazing the Liquid-$^{4}\mathrm{He}$ Surface*, Phys. Rev. Lett. **50**, 990 (1983).
[3] J. M. Doyle, J. C. Sandberg, I. A. Yu, C. L. Cesar, D. Kleppner, and T. J. Greytak, *Hydrogen in the Submillikelvin Regime: Sticking Probability on Superfluid $^{4}\mathrm{He}$*, Phys. Rev. Lett. **67**, 603 (1991).
[4] C. Carraro and M. W. Cole, *Role of Long-Range Forces in H Sticking to Liquid He*, Phys. Rev. B **45**, 12930 (1992).
[5] T. W. Hijmans, J. T. M. Walraven, and G. V. Shlyapnikov, *Influence of the Substrate on the Low-Temperature Limit of the Sticking Probability of Hydrogen Atoms on He Films*, Phys. Rev. B **45**, 2561 (1992).
[6] W. Kohn, *Quantum Mechanics of Sticking*, Surf. Rev. Lett. **01**, 129 (1994).
[7] F. Shimizu, *Specular Reflection of Very Slow Metastable Neon Atoms from a Solid Surface*, Phys. Rev. Lett. **86**, 987 (2001).
[8] F. Shimizu and J. Fujita, *Reflection-Type Hologram for Atoms*, Phys. Rev. Lett. **88**, 123201 (2002).
[9] G.-R. Wang, T. Xie, Y. Huang, and S.-L. Cong, *Quantum Reflection by the Casimir–Polder Potential: A Three-Parameter Model*, J. Phys. B: At. Mol. Opt. Phys. **46**, 185302 (2013).
[10] T. A. Pasquini, M. Saba, G.-B. Jo, Y. Shin, W. Ketterle, D. E. Pritchard, T. A. Savas, and N. Mulders, *Low Velocity Quantum Reflection of Bose-Einstein Condensates*, Phys. Rev. Lett. **97**, 093201 (2006).
[11] A. L. Marchant, T. P. Billam, M. M. H. Yu, A. Rakonjac, J. L. Helm, J. Polo, C. Weiss, S. A. Gardiner, and S. L. Cornish, *Quantum Reflection of Bright Solitary Matter Waves from a Narrow Attractive Potential*, Phys. Rev. A **93**, 021604 (2016).
[12] T. A. Pasquini, Y. Shin, C. Sanner, M. Saba, A. Schirotzek, D. E. Pritchard, and W. Ketterle, *Quantum Reflection from a Solid Surface at Normal Incidence*, Phys. Rev. Lett. **93**, 223201 (2004).
[13] A. Jurisch and H. Friedrich, *Realistic Model for a Quantum Reflection Trap*, Physics Letters A **349**, 230 (2006).
[14] A. Jurisch and J.-M. Rost, *Trapping Cold Atoms by Quantum Reflection*, Phys. Rev. A **77**, 043603 (2008).

[15] G. Rojas-Lorenzo, J. Rubayo-Soneira, S. Miret-Artés, and E. Pollak, *Quantum Reflection of Rare-Gas Atoms and Clusters from a Grating*, Phys. Rev. A **98**, 063604 (2018).
[16] B. Segev, R. Coté, and M. G. Raizen, *Quantum Reflection from an Atomic Mirror*, Phys. Rev. A **56**, R3350 (1997).
[17] F. Shimizu and J. Fujita, *Giant Quantum Reflection of Neon Atoms from a Ridged Silicon Surface*, J. Phys. Soc. Jpn. **71**, 5 (2002).
[18] S. Kallush, B. Segev, and R. Côté, *Manipulating Atoms and Molecules with Evanescent-Wave Mirrors*, Eur. Phys. J. D **35**, 3 (2005).
[19] A. Yu. Voronin, P. Froelich, and B. Zygelman, *Interaction of Ultracold Antihydrogen with a Conducting Wall*, Phys. Rev. A **72**, 062903 (2005).
[20] D. Kouznetsov, H. Oberst, A. Neumann, Y. Kuznetsova, K. Shimizu, J.-F. Bisson, K. Ueda, and S. R. J. Brueck, *Ridged Atomic Mirrors and Atomic Nanoscope*, J. Phys. B: At. Mol. Opt. Phys. **39**, 1605 (2006).
[21] R. Côté, E. J. Heller, and A. Dalgarno, *Quantum Suppression of Cold Atom Collisions*, Phys. Rev. A **53**, 234 (1996).
[22] A. R. Barnea, B. A. Stickler, O. Cheshnovsky, K. Hornberger, and U. Even, *Electrically Controlled Quantum Reflection*, Phys. Rev. A **95**, 043639 (2017).
[23] R. Côté and B. Segev, *Quantum Reflection Engineering: The Bichromatic Evanescent-Wave Mirror*, Phys. Rev. A **67**, 041604 (2003).
[24] M. Silvestre, T. P. Cysne, D. Szilard, F. A. Pinheiro, and C. Farina, *Tuning Quantum Reflection in Graphene with an External Magnetic Field*, Phys. Rev. A **100**, 033605 (2019).
[25] A. L. Harris, T. A. Saxton, and Z. G. Temple, *Recovery Time of Matter Airy Beams Using the Path Integral Quantum Trajectory Model*, Results in Physics **13**, 102253 (2019).
[26] J. Petersen, E. Pollak, and S. Miret-Artes, *Quantum Threshold Reflection Is Not a Consequence of a Region of the Long-Range Attractive Potential with Rapidly Varying de Broglie Wavelength*, Physical Review A **97**, (2018).